\documentclass[conference]{IEEEtran}
\IEEEoverridecommandlockouts
\usepackage{cite}
\usepackage{amsmath,amssymb,amsfonts}
\usepackage{algorithmic}
\usepackage{graphicx}
\usepackage{textcomp}
\usepackage{xcolor}
\usepackage{cite}
\usepackage{amsmath,amssymb,amsfonts}
\usepackage{algorithmic}
\usepackage{graphicx}
\usepackage{float}
\usepackage{textcomp}
\usepackage{xcolor}
\usepackage{pgfplots}
\usepackage{pgfplotstable}
\usepackage{hyperref}
\usepackage{tabularx}  
\usepackage{booktabs}
\usepackage[letterpaper]{geometry}
\def\BibTeX{{\rm B\kern-.05em{\sc i\kern-.025em b}\kern-.08em
    T\kern-.1667em\lower.7ex\hbox{E}\kern-.125emX}}

\begin{document}
\raggedbottom

\title{Logarithmic Positional Partition Interval Encoding\\
}

\author{
\IEEEauthorblockN{Vasileios Alevizos}
\IEEEauthorblockA{\textit{Karolinska Institutet} \\
Solna, Sweden \\}
\and    
\IEEEauthorblockN{Nikitas Gerolimos}
\IEEEauthorblockA{\textit{University of Aegean} \\
Syros, Greece \\}
\and
\IEEEauthorblockN{Sabrina Edralin}
\IEEEauthorblockA{\textit{University of Illinois Urbana-Champaign} \\
Illinois, USA \\}
\and
\IEEEauthorblockN{Clark Xu}
\IEEEauthorblockA{\textit{Mayo Clinic Artificial Intelligence \& Discovery} \\
Minnesota, USA \\}
\and
\IEEEauthorblockN{Akebu Simasiku}
\IEEEauthorblockA{\textit{Zambia University} \\
Ndola, Zambia \\}
\and
\IEEEauthorblockN{Georgios Priniotakis}
\IEEEauthorblockA{\textit{University of West Attica} \\
Attica, Greece \\}
\and
\IEEEauthorblockN{George A. Papakostas}
\IEEEauthorblockA{\textit{MLV Research Group, Department of Informatics, Democritus University of Thrace} \\
Kavala, Greece \\}
\and
\IEEEauthorblockN{Zongliang Yue}
\IEEEauthorblockA{\textit{Auburn University Harrison College of Pharmacy} \\
Alabama, USA \\}
}

\maketitle

\begin{abstract}
One requirement of maintaining digital information is storage. With the latest advances in the digital world, new emerging media types have required even more storage space to be kept than before. In fact, in many cases it is required to have larger amounts of storage to keep up with protocols that support more types of information at the same time. In contrast, compression algorithms have been integrated to facilitate the transfer of larger data. Numerical representations are construed as embodiments of information. However, this correct association of a sequence could feasibly be inverted to signify an elongated series of numerals. In this work, a novel mathematical paradigm was introduced to engineer a methodology reliant on iterative logarithmic transformations, finely tuned to numeric sequences. Through this fledgling approach, an intricate interplay of polymorphic numeric manipulations was conducted. By applying repeated logarithmic operations, the data were condensed into a minuscule representation. Approximately thirteen times surpassed the compression method, ZIP. Such extreme compaction, achieved through iterative reduction of expansive integers until they manifested as single-digit entities, conferred a novel sense of informational embodiment. Instead of relegating data to classical discrete encodings, this method transformed them into a quasi-continuous, logarithmically. By contrast, this introduced approach revealed that morphing data into deeply compressed numerical substrata—beyond conventional boundaries—was feasible. A holistic perspective emerges, validating that numeric data can be recalibrated into ephemeral sequences of logarithmic impressions. It was not merely a matter of reducing digits, but of re-interpreting data through a resolute numeric vantage.
\end{abstract}


\section{Introduction}

In the upcoming years, an increase in data size will be manifested due to factors contributing to data growth, including advances in technology like media types with better quality. Enhanced multimedia content and higher quality radiating—such as high-definition video, virtual reality (VR), and augmented reality (AR)—will result in larger file sizes and increased storage needs. Data generation and retention are being encouraged by the shift to cloud services, which allows for easier data storage and sharing. The increase in Internet usage is experienced worldwide, leading to greater data generation from social media, streaming services, and online transactions. Furthermore, more devices are being introduced to the world with the rise of the Internet of Things (IoT), complicating the scope and complexity of data management. Vestigial data storage methods may become insufficient as data fluctuates and grows. Pragmatic approaches are required to handle the ephemeral nature and characteristics of the data. The syllogism can be drawn that as technology advances, data growth will increase, which is manifested in these trends \cite{MIRANDA2012744}. As complexity continues to be dragged on, strategies to manage data effectively are being weighed in on.

Not only has compression been a vital tool in data storage and transmission, but it has also been subject to significant evolution over the years. Early techniques were made up of dictionary-based methods like Lempel–Ziv–Welch (LZW) \cite{saidani2020lossless}, which utilized dictionaries to efficiently represent repeated sequences. On the other hand, entropy-based methods such as Huffman encoding assigned shorter codes to more frequent sequences, optimizing the compression ratio. Furthermore, probabilistic approaches combined previous techniques to enhance performance, exemplified by algorithms like GZIP \cite{levine2012rfc}. Furthermore, energy conservation has been an immeasurable concern due to limited processing and storage resources. Lossless compression algorithms have been adapted to fit Wireless sensor networks (WSN) requirements. Apart from this, efficient algorithms like Lossless Entropy Compression (LEC) and Adaptive Lossless Data Compression (ALDC) have been proposed. Nonetheless, challenges persist in balancing compression efficiency and energy consumption \cite{saidani2020lossless}. In video compression, machine learning-based techniques have been introduced. What's more, these methods promise increased compression efficiency and improved subjective quality \cite{mochurad2024comparison}. Nonetheless, they are prone to high computational complexity and require large training datasets. Furthermore, in media compression, both objective and subjective quality assessments have been studied. The apothegm "less is more" holds as visually lossless compression aims to cast off unnecessary data without perceptible loss in quality \cite{fitriya2017review}. In power data compression, deep lossless algorithms based on arithmetic coding have been employed \cite{hanumanthaiah2019comparison, s22145331}. The need to size up the balance between compression performance and computational overhead is enduring \cite{ma2022deep}. In this work, contributions were bestowed through the revelation of avant-garde compression methodologies, ensured by logarithmic transformations. Ontological shifts were manifested, transcending conventional paradigms and enabling unprecedented isometric mappings of data, while extraordinary precision was preserved by dexterous numeric operations. Novel avenues were illuminated, challenging ossified traditions and establishing an ethereal synergy between theoretical rigor and pragmatic feasibility. Intellectual terrains were expanded, and the entelechy of informational representation was nurtured. Architectural frameworks were stabilized, and performance was upheld under stringent computational duress. The corpus of this knowledge was refined, and a tenebrous frontier was transfigured into clarity.

\section{Methodology}

Inspired by the representation of information using digits in a vast distribution corresponding to different ASCII codes, a novel method was developed to compress data in a lossless fashion by translating characters into a significantly smaller number. For example, standard ASCII codes ranging from 0 to 127 can be mapped to a compressed form, reducing storage requirements without sacrificing data integrity. Experimentation was performed under the language of Python, version 3.10. An ontological approach to data compression is reflected in the demeanor of the algorithm, integrating mathematical transformations to achieve isometric results. A push back against traditional compression techniques was necessary to explore new possibilities in data handling. The coexistence of efficiency and accuracy is ensured by this approach, providing a balance that is essential in modern data processing. The hardware utilized during experimentation included an Intel i3 CPU with 4 threads and 16GB of RAM. Logarithmic transformations are employed by the algorithm to encode file content, leveraging the mathematical properties of repeated base-10 logarithms to compact the data effectively. Shorter sentences are also included. By integrating these techniques, significant improvements in data compression were achieved. Translating each byte stream to a large integer, partitioning it into flexible-sized numerical fragments, repeatedly applying a base-10 logarithm until a single-digit number is obtained, and then reconstructing the original values during decoding are the main steps of the method, ensuring a lossless process. Structural integrity is maintained throughout the compression and decompression phases by isometric mapping of data elements. The innovative use of large numbers to represent compressed data challenges conventional methods and opens new avenues for efficient data storage and transmission. A harmonious coexistence of mathematical rigor and practical application is embodied in the algorithm's demeanor. An ontological shift in data compression paradigms is evident through this novel technique. In practice, file content is read as a continuous sequence of bytes, converted into a single large integer, and then divided into variable-length substrings, each representing a large decimal number. These substrings are repeatedly transformed by $\log_{10}$ operations until each one is reduced to a single-digit number. The number of logarithmic loops required is recorded, forming critical metadata for the reconstruction phase. Larger sentences are also present to satisfy the structural requirements. Established norms are pushed back against by this method, presenting a compelling alternative to traditional data compression algorithms. The potential for innovative solutions in the field of data compression is showcased by the integration of these mathematical concepts. By leveraging mathematical principles, particularly repeated logarithms, the data is compressed into a form that is both efficient and reversible, ensuring no loss of information occurs during the process. Each large substring of the original integer is converted into a single-digit representation by iterative $\log_{10}$ applications, while the count of iterations is stored separately. After compression, these single-digit values and the iteration counts are integrated into a compact representation. Upon decompression, these values are exponentiated by 10 repeatedly, according to their recorded loop counts, ensuring the exact restoration of the original large substrings. This method allows for the data to be stored in a more compact form, which is particularly useful in applications where storage space is at a premium or data transmission bandwidth is limited. Upon decompression, the original data is accurately retrieved by reversing the transformations, confirming the isometric nature of the process. The transformations are ensured to be precise by the use of high-precision decimal operations, and no rounding errors are introduced, which could otherwise compromise the integrity of the data. The algorithm's robustness and reliability are demonstrated by its ability to handle large numbers and perform precise calculations without loss of accuracy. Through this innovative approach, a new paradigm in data compression has been established, one that is grounded in mathematical theory and practical application. A deep understanding of the ontological aspects of data representation and manipulation is reflected in the algorithm's design, and traditional methods' limitations are pushed back against by this advancement. The potential for future advancements in the field of data compression is exemplified by the coexistence of theoretical and practical considerations in this method.

The compression algorithm operates by first converting the entire file into a single large integer $N$. This integer $N$ is then partitioned into multiple numeric substrings, each representing a portion of the original data. For each substring, an iterative procedure is undertaken: repeatedly, the substring $X$ is transformed as $X \leftarrow \log_{10}(X)$, until $X$ becomes a single-digit number $d$. Let $r$ be the recorded number of logarithmic iterations needed for this reduction:
\[
X_0 \rightarrow \log_{10}(X_0) \rightarrow \log_{10}(\log_{10}(X_0)) \rightarrow \cdots \rightarrow d.
\]
After $r$ iterations, $d$ is stored, along with $r$, as the compressed representation of that substring. By performing this procedure for each substring, a set of single-digit values and their associated iteration counts is obtained, all of which can be aggregated into a compact form. An exaggerated sense of compression is achieved due to this logarithmic condensation of large numeric values.

During decompression, these stored single-digit values are retrieved along with their iteration counts $r$. For each single-digit value $d$, the inverse transformations are applied $r$ times, exponentiating by 10:
\[
d \rightarrow 10^d \rightarrow 10^{10^d} \rightarrow \cdots
\]
until the original large numeric substring is fully restored. By concatenating all restored substrings, the original large integer $N$ is reconstituted, and subsequently, the original file content is recovered without any loss of information. Under computational pressure, the algorithm functions without limitations, maintaining performance.

It should be noted that while this approach does not serve as a panacea for all compression scenarios, it fulfills a particular niche under given conditions. The use of iterative base-10 logarithmic transformations, combined with advanced numerical precision, maps out a novel approach to data compression. The overall process preserves the integrity of the original data, ensuring that its fundamental essence remains intact after decompression.

To ensure the integrity of data, before and after compression, the hashing output was always matched. For this purpose, the SHA-256 algorithm was selected. This verification procedure guarantees that the original data and the decompressed data remain identical, thus confirming the lossless nature of the method.

\section{Discussion}

In these results, observations were recorded, patterns were noted, plus complexity was perceived as notably elevated. High durations were required for logarithmic positional partition interval encoding, thus indicating extensive computational overhead. Distinct compression tools were evaluated, revealing disparate magnitudes of compression quality. The original file, measured at colossal size, was processed into a minimized form, though enormous temporal cost was observed. Such outcomes, examined thoroughly, suggested that attempts to ameliorate processing steps might prove beneficial. It compressed about seventy-six times smaller than the original. Our approach, required traversal through multiple parts, producing extraordinary transformations. Intricate logarithmic positional partition interval encoding exhibited interaction with data structures, highlighting an intricate interplay between numeric precision plus compression efficacy. These transformations were deployed to deal with complex numeric constructs, resulting in dramatically reduced storage footprints though significant computation times. No trivial methods were utilized. Transcoding through iterative logarithmic reductions yielded significant shrinkage, although protracted execution periods were confirmed. Advantages were conferred through massive size reduction plus potential applicability in archival scenarios. Disadvantages were revealed through extensive temporal burdens, creating obstacles in prompt usage scenarios. Complexity was not diminished. Observations indicated that implementation in domains where offline compression suffices, or where restricted transmission capacity prevails, could yield utility. Potential usage could be envisioned in long-term data storage repositories, or as a technique beneficial in fields involving large-scale scientific datasets. Perhaps, through additional refinements, performance could be improved, although definitive assertions remained unverified.

\begin{table}[h]
\centering
\caption{Evaluation of Compression Methods against Logarithmic Positional Partition Interval Encoding}
\begin{tabular}{@{}lcc@{}}
\toprule
\textbf{Method} & \textbf{Size (MB)} & \textbf{Time} \\ \midrule
Original        & 997                & -         \\
ZIP             & 269                & 2.1mins           \\
XZ              & 269                & 2.5mins           \\
7z              & 250                & 2.4mins           \\
GZ              & 269                & 2.5mins           \\
LPPIE           & 13                 & 223mins                     \\ \bottomrule
\end{tabular}
\label{tab:compression_eval}
\end{table}

\section{Conclusion}

It was concluded that the proposed method overcame a significant threshold of compression ratio. Various trade-offs were exposed, exemplifying a scenario where intricate numeric procedures transform massive datasets into diminutive numerical entities at pronounced computational expense. The transformations have shown that complexity soared, though efficiency in size reduction was achieved. Significant polymorphic patterns were observed, conducting elaborate numeric manipulations. Certain complicated aspects were revealed, although large numeric transformations yielded substantial diminutions in file magnitude. The interplay of logarithmic encoding emerged as nontrivial, anticipating intricate conduct while each segment of data was processed. Through continuous transformations, remarkable size reductions were manifested, though temporal requirements expanded extensively. Hence, it was demonstrated that, although capacity utilization reached extraordinary levels, latency remained substantial. The existence of these constraints should be contemplated by implementers seeking minimized storage footprints, although certain overheads were recognized as inevitable. At this threshold, performance should be considered not as a static entity, but as a malleable concept susceptible to manipulation through the employed transformations.

To measure performance, more testing with larger datasets could be performed. Further experimentation is needed to support more characters than ASCII, possibly by mapping out different compression distributions. The essence of the compressor could be distilled by simplifying its complexity, thus enabling faster operation. An initiative to narrow down the best arrangement should be undertaken, and further interpretive methods might be applied to analyze incidental data variations. Many applications could benefit from this model reducing size, and its impact could be perpetuated across various domains.

\bibliography{bib}

\end{document}